\documentclass{article}
\usepackage[dvips]{graphicx}
\newtheorem{proposition}{Proposition}
\def\tr{{\rm tr}}

\begin{document}

\title{Isotropization in the approach to big rip singularities for
Cardassian models}

\author{Nikolaus Berndt\\
Institut f\"ur Physik\\
Humboldt-Universit\"at zu Berlin\\
Newtonstrasse 15\\
12489 Berlin, Germany\\
\\
and\\
\\
Alan D. Rendall\\
Max-Planck-Institut f\"ur Gravitationsphysik\\
Albert-Einstein-Institut\\
Am M\"uhlenberg 1\\
14476 Potsdam, Germany}

\date{}

\maketitle

\begin{abstract}
Cardassian models are an alternative to general relativity which have
been proposed as an approach to explaining accelerated cosmic 
expansion while avoiding directly introducing dark energy. They are
generally formulated only in the homogeneous and isotropic case.
In this paper an extension of the usual formulation to general 
spatially homogeneous geometries is given. A characteristic feature of 
many classes of Cardassian models is the occurrence of big rip 
singularities where the scale factor tends to infinity after a finite
time. It is shown that big rip singularities are also widespread in
more general homogeneous cases. It is also shown that there is 
isotropization in the approach to a big rip singularity which bears
a strong resemblance to the late-time isotropization observed in 
cosmological spacetimes which accelerate forever in the future.
\end{abstract}

\section{Introduction}

The realization, based on observational data, that the expansion of the 
universe is accelerated has led to a lot of work which modifies 
traditional cosmological models by introducing exotic types of matter  
or replacing general relativity by something else. The focus of interest
in the following is an example of the second approach. It concerns the
Cardassian models which were first introduced in \cite{freese02}. The 
standard solutions of the Einstein equations applied in cosmology are 
those which are homogeneous and isotropic. These are the 
Friedmann-Lemaitre-Robertson-Walker (FLRW) models. Often attention is
concentrated on the spatially flat case, where the spacetime metric 
takes the form 
\begin{equation}
-dt^2+a^2(t) (dx^2+dy^2+dz^2)
\end{equation}
with scale factor $a(t)$. In these solutions a central role is 
played by the Friedmann equation
\begin{equation}\label{friedmann}
\left(\frac{\dot a}{a}\right)^2=\frac{8\pi\rho}3+\frac{\Lambda}3
\end{equation}
where $\rho$ is the energy density, $\Lambda$ is the cosmological constant 
and geometrized units are used where the gravitational constant and the 
speed of light take the numerical value unity. The other independent 
Einstein equation is the evolution equation
\begin{equation}\label{evolution}
\frac{\ddot a}{a}=-\frac{4\pi}{3}(\rho+3p)+\frac{\Lambda}3
\end{equation}
where $p$ is the pressure. As a consequence of energy-momentum conservation
the matter quantities $\rho$ and $p$ satisfy
\begin{equation}\label{euler}
\dot\rho+3H(\rho+p)=0
\end{equation}
where $H=\dot a/a$ is the Hubble parameter. It is well known that if 
equation (\ref{friedmann}) is satisfied at some time $t_0$ and equations 
(\ref{evolution}) and (\ref{euler}) are satisfied at all times then 
(\ref{friedmann}) is satisfied at all times. This is a special case of the 
general phenomenon of propagation of the constraints in solutions of the 
Einstein equations. Accelerated expansion means that $\ddot a>0$ and it 
follows from (\ref{evolution}) that this is only possible if $\Lambda>0$ or 
$\rho+3p<0$. Many matter models in general relativity satisfy the strong 
energy condition $\rho+3p\ge 0$ which corresponds to the fact that gravity
is attractive. Accelerated expansion 
is only possible in general relativity if either $\Lambda>0$ or there is
a matter field (dark energy) which violates the strong energy condition.
The idea of \cite{freese02} is to replace general relativity by another 
theory so as to make accelerated cosmological expansion possible with
a vanishing cosmological constant and without the need for matter violating
the strong energy condition.

In the Cardassian models of \cite{freese02}, which are formulated for 
homogeneous and isotropic geometries, the Friedmann equation 
(\ref{friedmann}) is replaced by the equation 
\begin{equation}\label{friedmannc}
\left(\frac{\dot a}{a}\right)^2=\frac{8\pi}3(\rho+f(\rho))
\end{equation}
for a smooth function $f$. The standard choice for $f(\rho)$ has the 
power-law form $B\rho^n$ for constants $B$ and $n$ but in principle the
formalism can be developed for any function $f$. In later work 
\cite{gondolo02} this approach was changed and the Cardassian term
is only a function of part of the energy density of the matter, the
rest energy density. This splitting requires the input of more information 
than the energy-momentum tensor alone. The analysis in this paper will 
concentrate on the original version although the later version will be
commented on. 

In \cite{freese02} a possible origin of the additional term in the 
Cardassian framework was discussed in the context of braneworld models. 
A Cardassian term also arises in loop quantum cosmology as a manifestation 
of the discreteness of spacetime \cite{singh06}. In the latter case the 
constant $B$ is negative.

If the Cardassian models are to be regarded as defining a theory of
gravity then there should exist a definition of these models applying
to solutions without symmetry which reduces to the known definition in
the homogeneous and isotropic case. A general theory of this type is
also of practical importance for comparisons with cosmological 
observations. Some of these comparisons use linearization around the
homogeneous and isotropic models. In order to obtain the field equations
for linearized perturbations the natural procedure is to start with a 
general theory and linearize. To the authors' knowledge no such general
theory has been found.  The comparison of theory with some aspects of
observations, e.g. estimates of the age of the universe or supernova
data, only requires the scale factor but other data such as that on 
cosmic microwave background fluctuations requires more structure 
(including an understanding of linear perturbations) on the theoretical side 
to allow a comparison. The fact that calculations could be done for 
linearized perturbations is based on the use of what will be called the
'effective fluid' in what follows. 

Because of the high symmetry of 
homogeneous and isotropic geometries any Cardassian model with a specific
choice of $f$ and matter modelled by a perfect fluid with a specific 
choice of equation of state $p=g(\rho)$ is equivalent to general relativity
with an effective fluid with equation of state $p=\tilde g (\rho)$, where
$\tilde g$ is determined uniquely by $f$ and $g$. 
Starting from this observation it is possible to
use the linearized Einstein-Euler equations with this effective equation
of state to do perturbation theory. If, on the other hand, another specific
description of matter, such as kinetic theory, is used then things are more
complicated. For a particular solution of the Einstein equations coupled
to this matter model which is homogeneous and isotropic it is possible to
define energy density and pressure and, with a suitable monotonicity
assumption, an equation of state relating the two. However this 'equation
of state' depends on the solution (and hence on the initial data) and
as a consequence not only on the choice of matter model. Thus it cannot be 
used to define a generalization to the inhomogeneous case. The unfortunate 
lack of a general theory cannot simply be overcome by an appeal to braneworld 
models or loop quantum cosmology since these constructions have also only been 
carried out fully under symmetry assumptions. 

It will be shown that it is possible to define a natural generalization of 
the Cardassian models to spacetimes which are homogeneous but not necessarily 
isotropic. This accomodates any form of matter. For general matter models
the energy-momentum tensor is intrinsically anisotropic - it has distinct 
principal pressures - and so this cannot be replaced by any effective perfect 
fluid description. A limitation 
of this definition is that, as will be shown in section 3, for the most 
general classes of homogeneous geometries and matter it requires the 
assumption that the cosmological expansion never vanishes in order to have
a regular evolution equation. An interesting dynamical feature of Cardassian 
models in contrast to general relativity with conventional fluid is the 
occurrence of big rip singularities \cite{caldwell03}. This means that 
the scale factor $a(t)$ tends to infinity in a finite proper time. The main 
result of this paper is that under general assumptions homogeneous 
spacetimes isotropize in the approach to a big rip singularity. This is
analogous to the cosmic no hair theorem for the late-time behaviour of
solutions of the Einstein equations with positive cosmological constant
and its proof in the homogeneous case due to Wald \cite{wald83}.     

The structure of the paper is as follows. Section 2 contains an analysis 
of the dynamics of homogeneous and isotropic Cardassian models using an 
effective potential. The method of an effective fluid is also described, 
together with its limitations. In section 3 the generalization to cases
which are homogeneous but not isotropic is presented and isotropization 
near a big rip singularity is proved. In the last section possible further 
developments of Cardassian models and related topics are discussed. This
paper is based in part on the diploma thesis of the first author 
\cite{berndt}. 

\section{Isotropic models}

The basic equation for Cardassian models is (\ref{friedmannc}). The analysis
of this section is restricted to isotropic models but spatial curvature is
included so that (\ref{friedmannc}) becomes
\begin{equation}\label{friedmannck}
\left(\frac{\dot a}{a}\right)^2=\frac{8\pi}3(\rho+f(\rho)).
-\frac{k}{a^2}
\end{equation}
The curvature parameter $k$ is $+1$, $0$ or $-1$. The 
equation (\ref{euler}) expressing energy-momentum conservation remains 
unchanged. Differentiating (\ref{friedmannck}) with repect to time and
using (\ref{euler}) in the result gives the following evolution equation:
\begin{equation}\label{evolutionc}
\frac{\ddot a}{a}=-\frac{4\pi}{3}(\rho+3p)+\frac{8\pi}3\left[f(\rho)
-\frac32 (\rho+p)f'(\rho)\right].
\end{equation}
It is clear from this expression that accelerated expansion is possible 
even when the matter satisfies the strong energy condition. These equations 
for describing gravity can in principle be combined with any model of matter. 
The equations of motion of the matter are assumed to take exactly the same
form as in general relativity. The remainder of this section will concentrate 
on the case where the matter is described by a perfect fluid with equation of 
state $p=g(\rho)$. Other matter models will be included in the discussion of 
anisotropic geometries in the next section. 

The dynamics of isotropic models with a fluid will now 
be analysed. A convenient way of doing this is to use the analogy of the
motion of a particle in a potential. It is assumed that the function $g$
defining the equation of state is non-negative with $g(0)=0$ and satisfies 
$0\le g'(\rho)\le 1$ for all values of $\rho$. Define a quantity $r$
representing the mass density of the fluid by
\begin{equation}\label{defr}
r(\rho)=\exp\left\{\int_1^\rho (\sigma+g(\sigma))^{-1}d\sigma\right\}
\end{equation}
Under the given assumptions $0<dr/d\rho\le 1$. From (\ref{defr}) it can
be concluded that $r\le\rho\le r^2$ for $\rho\ge 1$ and $r^2\le\rho\le r$
for $\rho\le 1$. It follows that the mapping from $\rho$ to $r$ can be 
inverted. Integrating equation (\ref{euler}) shows that $a^3 r$ is time 
independent. Combining this with
the conservation law results in a relation $\rho=h(a)$ for a function
$h$ depending on $g$. In the particular case that $g$ is a linear
function of $\rho$ the function $h$ is equal to a constant times a 
power of $a$. Define a potential $V$ by 
$V(a)=-\frac{4\pi}3a^2(h(a)+f(h(a)))$. Then 
the dynamics of $a$ is equivalent to that of a particle in this
potential. The particle has energy $-k/2$ as a consequence of 
(\ref{friedmannck}). The picture with the potential is useful for giving 
an intuitive method for seeing how the dynamics looks. It can however
do more and give rigorous proofs of the main features of the dynamics.
For the convenience of the reader this will be shown in detail in an
appendix.

Consider the case, which may be called the standard case, where the
Cardassian term is defined by $f(\rho)=B\rho^n$ and the equation of
state of the perfect fluid is $g(\rho)=(\gamma-1)\rho$ with 
$1\le\gamma\le 2$. Then $\rho=Aa^{-3\gamma}$ for a positive constant 
$A$ and 
\begin{equation}\label{potential1}
V(a)=-\frac{4\pi}3(Aa^{-3\gamma+2}+BA^na^{-3n\gamma+2}). 
\end{equation}
The cases $n=0$ and $n=1$ will be excluded from consideration since the 
first can be identified with general relativity with cosmological constant 
and the second is general relativity with vanishing cosmological constant in
disguise. These cases are already sufficiently well known. The domain of 
definition of the potential
is $(0,\infty)$ since $A$ is positive and $-3\gamma+2$ is negative. The 
derivative $V'(a)$ has a zero $a_*$ if and only if either $B>0$ and 
$n<\frac2{3\gamma}$ or $B<0$ and $n>\frac2{3\gamma}$. In both cases
\begin{equation}
a_*=\left[\frac{BA^{n-1}(2-3n\gamma)}{3\gamma-2}
\right]^{\frac{1}{3(n-1)\gamma}}.
\end{equation}
Important information about the potential is provided by the sign of
$V''$ at the critical point. 
\begin{equation}
V''(a_*)=-\frac{4\pi}3a_*^{-3\gamma}A[(-3\gamma+2)(-3\gamma+1)
+BA^{n-1}(-3n\gamma+2)(-3n\gamma+1)a_*^{-3(n-1)\gamma}]
\end{equation}
From this it follows that for $n>1$ the potential has a minimum at $a_*$ 
while for $n<1$ it has a maximum there. It is now easy to read off 
information about the qualitative behaviour of the solution using 
the results of the Appendix. When doing this Figs. \ref{cardtwofig}
-\ref{cardfig2} are useful for the purposes of orientation. In each of
these figures a representative potential is shown for the given range 
of parameters.
\begin{figure}
 \begin{center}
 \includegraphics[width=.6\textwidth]{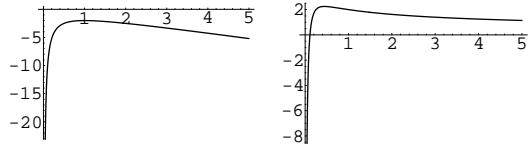}
 \end{center}
 \caption{The cases $B>0, n<\frac{2}{3\gamma}$ and 
$B<0, \frac{2}{3\gamma}<n<1$.}
\label{cardtwofig}
\end{figure}
\begin{figure}
 \begin{center}
 \includegraphics[width=.6\textwidth]{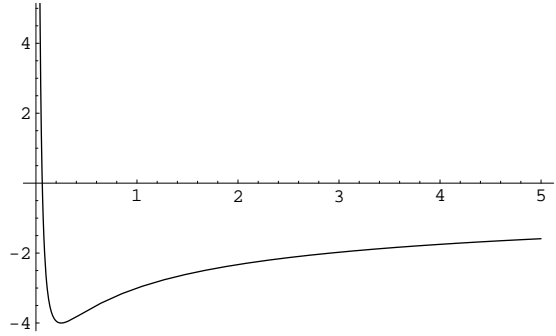}
 \end{center}
 \caption{The case $B<0, n>1$.}
\label{cardfig2}
\end{figure}
Since the universe is expanding today only
those solutions will be considered which are expanding at some time. Note 
that this excludes the exceptional time-independent solutions which occur
when $B>0$, $n<\frac2{3\gamma}$ or $B<0$, $n>\frac2{3\gamma}$ and $k=V(a_*)$. 
The results will be summarized in a proposition.

\begin{proposition}\label{mainprop} 
Let $a(t)$ be the scale factor of an isotropic and 
homogeneous solution of a 
Cardassian model with the matter being a perfect fluid with linear equation
of state $p=(\gamma -1)\rho$ and $1\le\gamma\le 2$. Suppose that there is
a time $t_0$ with $\dot a(t_0)>0$. Then if neither $B$ nor $n-\frac2{3\gamma}$ 
is zero the following mutually exclusive cases occur:
\begin{enumerate}
\item{The solution is periodic; occurs for all solutions with $B<0$, 
$n>1$ and $k>0$.}
\item{$a$ starts from zero at a finite time, increases to a maximum
and returns to zero in finite time; occurs for all solutions with 
$B>0$, $n<\frac2{3\gamma}$, $V(a_*)>-k/2$ and $a<a_*$, all solutions with 
$B>0$, $n>\frac2{3\gamma}$ and $k>0$, all solutions with $B<0$, 
$\frac2{3\gamma}<n<1$, $V(a_*)>-k/2$ and $a<a_*$ and all solutions with 
$B<0$ and $n<\frac2{3\gamma}$}
\item{$a$ tends to $\infty$ in the distant past, is decreasing before a 
certain time, reaches a minimum value and is increasing thereafter; occurs 
for all solutions with $B>0$, $n<\frac2{3\gamma}$, 
$-k/2<V(a_*)$ and $a>a_*$, all solutions with $B<0$,  $\frac2{3\gamma}<n<1$, 
$-k/2<V(a_*)$ and $a>a_*$ and all solutions with $B<0$, $n>1$ and $k<0$}
\item{$a$ increases in a monotone way from zero at a finite time; 
all solutions with $B>0$, $n<\frac2{3\gamma}$ and $-k/2\ge V(a_*)$, all 
solutions with $B>0$, $\frac2{3\gamma}<n$ and $k<0$ and all solutions with 
$B<0$, $\frac2{3\gamma}<n<1$ and $-k/2\ge V(a_*)$}. If $-k/2>V(a_*)$ then 
$a\to\infty$ as $t\to\infty$ while if $-k/2=V(a_*)$ the solution approaches 
$a_*$ as $t\to\infty$.
\end{enumerate}
\end{proposition}

The case $B=0$ corresponds to the well-known case of general relativity.
If $n=\frac2{3\gamma}$ the solutions behave as in the second or fourth items 
in the list of the proposition, depending on the value of 
$\frac{4\pi AB^{n-1}}3$ in comparison to $-k/2$. They are equivalent to models
without Cardassian term and with a modified value of the curvature.
One of the most interesting types of
behaviour included in the statement of the proposition is the fact
that for certain cases the big bang singularity is replaced by a bounce. It 
is this which, in particular, makes oscillating solutions possible.

Next more detail will be obtained about the leading order asymptotics
of the scale factor as $a$ approaches zero or infinity. Without loss of
generality it may be assumed that the approach of $a$ to zero takes place 
in the past and its approach to infinity takes place in the future.
Consider first the limit $a\to 0$. Let the degenerate case when $n=1$ and 
$B=-1$ be excluded from consideration. For $a$ small 
$V(a)=-\frac12 C^2 a^{-2p}(1+o(1))$ as $a\to 0$ for constants $C$ and $p$ with 
$p\ge 1/2$. When $a\to 0$ the curvature parameter is negligible compared to 
the potential and so $\dot a=Ca^{-p} (1+o(1))$. Moreover one of the terms
in the potential dominates the other. Hence 
\begin{equation}
a=(C(p+1))^{\frac1{p+1}}(t-t_0)^{\frac1{p+1}}(1+o(1))
\end{equation}
where $t_0$ is the time at which $a$ vanishes. If $n>1$ the Cardassian
term dominates as $a\to 0$ and putting in the relevant constants gives
\begin{equation}\label{inflation}
a=(12\pi Bn^2\gamma^2A^n)^{\frac{1}{3n\gamma}}
(t-t_0)^{\frac{2}{3n\gamma}}(1+o(1)).
\end{equation}
If $n<1$ the Cardassian term is negligible for $a$ small and 
\begin{equation}
a=(12\pi \gamma^2 A)^{\frac{1}{3\gamma}}
(t-t_0)^{\frac{2}{3\gamma}}(1+o(1)).
\end{equation}

At a big rip singularity, where $a$ tends to infinity at a finite time
$t_*$ it is again the case that the curvature parameter becomes 
negligible and one of the terms in the potential dominates the other.
In this case $\dot a=Ca^q (1+o(1))$ with $q>0$ and $C>0$. For
a finite-time singularity it is necessary that $q>1$. This corresponds
to $n<0$. Letting $b=a^{-1}$ and reversing the direction of time 
recovers the situation analysed above, with a weaker restriction on
the powers. In fact $q=1$ corresponds to $p=1$. Putting in the relevant
constants allows the following asymptotics to be derived using the
previous analysis:
\begin{equation}
a=(12\pi Bn^2\gamma^2 A^n)^{\frac{1}{3n\gamma}}
(t_*-t)^{\frac{2}{3n\gamma}}(1+o(1)).
\end{equation}
Here $t_*$ is the time at which the singularity occurs. With some more
work the $o(1)$ error terms in these expressions can be replaced with 
specific positive powers of $t-t_0$ or $t_*-t$ repectively \cite{berndt}.

Now the effective fluid approach to isotropic Cardassian
models will be described. The idea is to define a new energy density by
$\tilde\rho=\rho+f(\rho)$ and a new pressure $\tilde p$ so that 
$d\tilde\rho/dt=-\frac{3\dot a}{a}(\tilde\rho+\tilde p)$. If this can be
done and if the expression for $\tilde\rho$ in terms of $\rho$ can be 
inverted a relation of the form $\tilde p=\tilde g (\tilde\rho)$ can be
obtained. Then the dynamical properties of the Cardassian model are
equivalent to that of the Einstein equations coupled to a perfect fluid
with equation of state $\tilde p=\tilde g (\tilde\rho)$. It should,
however, be noted that the effective fluid has exotic properties such
as violating the strong and weak energy conditions in general, the latter
in the case $B<0$. In this construction the relation
\begin{equation}
\tilde p=p+f'(\rho)(\rho+p)-f(\rho)
\end{equation}
is obtained. To solve for $\rho$ in terms of $\tilde\rho$ it suffices
for the condition $f'(\rho)>-1$ to be satisfied since then the inverse
function theorem can be applied. In the case of a power-law
Cardassian term this works in the case $Bn>0$ but not when $Bn<0$.
In the latter case the attempt to carry out this construction leads to
an equation of state which is not single-valued. Thus this transformation
cannot be carried out globally for $Bn<0$. Note that in the case of forever
expanding models for which $\rho\to 0$ as $t\to\infty$ the transformation can
be carried out locally so as to apply to solutions at late times. When
$f(\rho)=B\rho^n$ and $p=(\gamma-1)\rho$ the equation of state for the 
effective fluid approaches the linear form $\tilde p=(n\gamma-1)\tilde\rho$
as $\tilde\rho\to 0$. By comparison with models with a fluid with exactly
linear equation of state it may be guessed that power-law inflation will
occur for $n$ in the range $(\frac2{3\gamma},1)$. This is confirmed by
the formula (\ref{inflation}).

To end this section the alternative approach to Cardassian models will
be considered briefly. In this case the expression $\rho+f(\rho)$ in 
(\ref{friedmannc}) is replaced by $\rho_1+f(\rho_1)+\rho_2$ for two 
independent quantities $\rho_1$ and $\rho_2$. Here $\rho_1$ represents
the rest energy density and $\rho_2$ all other forms of energy contained
in the matter. The equations of motion for the matter are 
$d\rho_1/dt=-3H\rho_1$ and $d\rho_2/dt=-3H(\rho_2+p)$.
In the case that $p=(\gamma-1)\rho$ the dynamics can be modelled by the 
potential
\begin{equation}\label{potential2}
V(a)=-\frac{4\pi}3(A_1a^{-1}+BA_1^na^{-3n+2})+A_2a^{-3\gamma+2}. 
\end{equation}
The critical points of this potential are not so easy to determine as those
of the potential in (\ref{potential1}) and it may not be possible to 
compute them explicitly. Nevertheless it should be possible to do a 
complete analysis of the qualitative features of the dynamics with some
more work.

\section{Isotropization near big rip singularities}

There is a natural way to generalize the Cardassian models to the case of 
spacetimes which are homogeneous but not isotropic. The basic principle
is to require the propagation of the constraint equations. The 
Hamiltonian constraint, which corresponds to the Friedmann equation
(\ref{friedmann}) in the isotropic case, is modified when passing from
general relativity to the Cardassian case in exactly the same way as was
done for isotropic models
- the energy density $\rho$ is replaced by $\rho+f(\rho)$ and all other
terms are left unchanged. The momentum constraint is left as it is.
The equations of motion of the matter, and hence the equation for 
energy-momentum conservation, are as in general relativity. 
Then the question of the propagation of the constraints is posed.
If both the constraints are satisfied at some time $t=t_0$ and a 
certain evolution equation generalizing (\ref{evolutionc}) is satisfied
at all times together with the equations of motion of the matter are the 
constraints satisfied at all times? It turns out that 
there is an evolution equation which gives a positive answer to
this question and reduces to the usual evolution equation for general
relativity when the Cardassian term $f$ vanishes. This equation will now 
be written down. 

The spatially homogeneous spacetimes can be divided into the Bianchi
spacetimes of types I-IX and the Kantowski-Sachs spacetimes. In the 
following only Bianchi models will be considered, although it would
presumably be possible to carry out a similar analysis in the 
Kantowski-Sachs case. In the case of the Bianchi models it is possible 
to assume without loss of generality that the spatial manifold is 
simply connected. For passing from a Bianchi model to its universal cover
does not change the dynamics. Once this has been done the spatial manifold
can be identified with a simply connected three-dimensional Lie group
$G$ and the metric and other geometrical objects can usefully be 
parametrized by their components in a basis of left-invariant one-forms
$\theta^i$ on $G$ with a dual basis $e_i$ of vector fields. The spatial 
metric is specified by a matrix $g_{ij}(t)$ depending only on time.
The spacetime metric is of the form $-dt^2+g_{ij}\theta^i\theta^j$.
Let $k_{ij}$ be the second fundamental form of the hypersurfaces of
constant $t$ and define $H=-\frac13\tr k$. In the isotropic case this
quantity $H$ reduces to the Hubble parameter introduced previously.
 
The modified Hamiltonian constraint is 
\begin{equation}\label{hamiltonian}
R-k_{ij}k^{ij}+(\tr k)^2=16\pi (\rho+f(\rho)).
\end{equation}
The evolution equation for the second fundamental form is
\begin{equation}\label{kevol}
\frac{dk^i{}_j}{dt}=R^i{}_j+(\tr k)k^i{}_j-8\pi(S^i{}_j-\frac12\tr S\delta^i_j)
-4\pi\rho\delta^i_j+U^i{}_j
\end{equation}
with trace
\begin{equation}\label{trkevol}
\frac{d(\tr k)}{dt}=R+(\tr k)^2+4\pi\tr S-12\pi\rho+\tr U.
\end{equation}
Combining this with the generalized Hamiltonian constraint gives
\begin{equation}\label{trkevol2}
\frac{d(\tr k)}{dt}=\frac13 (\tr k)^2+4\pi(\rho+\tr S)+\tr U+16\pi f(\rho).
\end{equation}
Here $S_{ij}=T_{ij}$ is the spatial projection of the energy-momentum tensor
and 
\begin{equation}\label{udef}
U^i{}_j=4\pi\left[ -2f(\rho)+f'(\rho)\left(\rho
-\frac{\nabla_l j^l}{\tr k}+\frac{k^l{}_mS^m{}_l}{\tr k}\right)\right]
\delta^i_j.
\end{equation}
In the isotropic case (\ref{trkevol2}) and (\ref{udef}) reduce to 
(\ref{evolutionc}). Note that $U^i{}_j$ is in general only well defined as 
long as $\tr k\ne 0$ since there are two terms with $\tr k$ in the 
denominator. The first of these vanishes in Bianchi Class A while the second 
is regular in the case of an untilted perfect fluid. (For the definitions of 
these terms the reader is referred to \cite{wainwright}.) Without these 
restrictions it could happen that the evolution breaks down. It is however
the case that for the types of solutions which will be of most interest
in the following some control of the apparently singular terms can be 
obtained. The source of this control are some inequalities implied by the
Hamiltonian constraint which will now be presented. Equation
(\ref{hamiltonian}) can be rewritten as
\begin{equation}\label{hamiltonian2}
\frac23 (\tr k)^2=\sigma_{ij}\sigma^{ij}-R+16\pi (\rho+f(\rho))
\end{equation}
where $\sigma_{ij}$ is the tracefree part of $k_{ij}$. In Bianchi types
I-VIII it is known that $R\le 0$ \cite{wald83}. Thus if a solution is of one 
of these Bianchi types and if $f$ is non-negative then if $\tr k$ tends to 
zero at some time it follows that $\rho$ tends to zero, with 
$\rho\le \frac{1}{24\pi} (\tr k)^2$,  $f(\rho)\le \frac1{24\pi} (\tr k)^2$
and $\sigma_{ij}\sigma^{ij}\le \frac23 (\tr k)^2$.

In spacetimes which are not isotropic the momentum constraint becomes
non-trivial and must be taken into account. As has already been mentioned,
for the Cardassian models it is assumed that the momentum constraint takes 
exactly the same form as in general relativity. This assumption is used in 
the calculation to verify that the Hamiltonian constraint propagates. In 
showing that the momentum constraint itself propagates the key feature of the 
modified evolution equation which is used is that the modification only 
affects the trace while the tracefree part of the equation remains unchanged.

It will now be investigated how solutions of these equations behave near 
a big rip singularity. This means by definition that the determinant
$\det g$ of the spatial metric or equivalently, in more geometrical language, 
its volume form tends to infinity in finite time. Assumptions 
will be made which are analogous to those made in \cite{wald83}.
Consider a spacetime of Bianchi type I-VIII which is expanding at some 
time and satisfies the Cardassian equations introduced above. Suppose 
further that the matter content of spacetime satisfies the dominant and
strong energy conditions. Thus there is no dark energy included. Finally,
suppose that the Cardassian term is of the form $f(\rho)=B\rho^n$ with 
$B>0$ and $n<0$. 

It will now be shown that in a spacetime of this type $\tr k$ remains 
negative as long as a smooth solution exists. In fact it will be
shown that as long as $\tr k$ is bounded there cannot exist a sequence
of times $t_n$ approaching the end of an interval where the solution
exists for which $\tr k (t_n)\to 0$. For this purpose, suppose that there 
is a smooth solution defined on a time interval $[t_0,T)$ with $\tr k$ 
bounded and a sequence $t_n\to T$ with the property in question. 
Let $Y$ denote the right hand side of (\ref{trkevol2}). Then
\begin{equation}\label{fund1}
\tr k (t)=\tr k (t_n)-\int_t^{t_n}Y(s)ds
\end{equation}
Using the remarks following (\ref{hamiltonian2}) it can be seen that all
terms in $Y$ except $\tr U$ can be bounded by an expression of the form
$C (\tr k)^2$ for a constant $C$. Under the assumption that $\tr k$ is 
bounded this implies that $\tr U\le C|\tr k|$ for a constant $C$.
The terms in $\tr U$ involving $f$ 
satisfy the same type of bound. It remains to consider the expressions
in $Y$ where $\tr k$ occurs in the denominator. In order to obtain an estimate 
for $\nabla_i j^i$ it is useful to evaluate it in an orthonormal frame. There 
the components of $j$ can be bounded by $\rho$ due to the dominant energy 
condition while the rotation coefficients are of order one. Thus 
$|\nabla_i j^i|\le C\rho$ for a constant $C$ and 
$|\nabla_i j^i|/(\tr k)\le C|\tr k|$.
To deal with the last term in $\tr U$ first note that 
\begin{equation}\label{decomp}
k^l{}_m/(\tr k)= \sigma^l{}_m/(\tr k)+\frac13\delta^l_m.
\end{equation} 
Thus there is an explicit contribution $\frac{4\pi}3 f'(\rho)\tr S$ to $\tr U$.
This is no more difficult to estimate than the terms already discussed. For 
the remaining quantity to be estimated it is possible to use the inequality 
that $\sigma^i{}_jS^j{}_i\le (S^i_jS^j_i)^{1/2}(\sigma^i{}_j\sigma^j{}_i)^{1/2}$.
The dominant energy condition shows that any component of $S$ in an 
orthonormal frame is bounded by $\rho$. Hence the last expression can 
be bounded by $C\rho$. Putting all this information together and writing
$\tau=t_n-t$ gives an inequality of the form
\begin{equation}\label{gronwall}
|\tr k| (\tau)\le |\tr k(0)| +C\int_0^\tau |\tr k(s)| ds.
\end{equation} 
Applying Gronwall's inequality \cite{hartman} and letting $n\to\infty$ gives
the conclusion that $\tr k=0$
on the whole interval, in contradiction to the orginal assumptions. This
proves the desired result concerning $\tr k$. Note that it follows form
this result that $|\tr k|$ has a strictly positive lower bound on the 
given interval. 

Estimates for the spacetime will be obtained with 
the help of a quantity $Z$. This is modelled on the quantity $S$
introduced for similar purposes in \cite{moss} which in turn was influenced
by the arguments of \cite{wald83}. 
\begin{equation}
Z=3H^2-8\pi f(\rho)=8\pi\rho+\frac12\sigma^{ij}\sigma_{ij}-\frac12 R.
\end{equation}
Here the equality of the second and third expressions follows from the 
modified Hamiltonian constraint (\ref{hamiltonian}). The time derivative 
of $Z$ is given by
\begin{equation}\label{dzbydt}
\frac{dZ}{dt}=-2HZ-2H\left[\sigma_{ij}\sigma^{ij}
+4\pi(\rho+\tr S)\right]\le -2HZ.
\end{equation}
Since it has been proved that $H>0$ it can in particular be concluded that
$Z$ is non-increasing.
Now introduce a quantity $l(t)$ which is positive and satisfies $\dot l/l=H$. 
This is a replacement for $a$ in the isotropic case. At a big rip singularity 
$l$ tends to infinity. It follows from (\ref{dzbydt}) that
\begin{equation}
\frac{d}{dt}(\log Z)\le -2\frac{d}{dt}(\log l).
\end{equation} 
Hence $d/dt (\log (Z l^2))\le 0$. As a consequence $Z\le C_1l^{-2}$ for 
a positive constant $C_1$. It follows that $\rho\le (C_1/8\pi)l^{-2}$.
Now this can be put back into the modified Hamiltonian constraint to
obtain a useful lower bound for $\dot l/l$, using the fact that in Bianchi 
types I-VIII the scalar curvature $R$ is non-positive. Compare the remarks 
following equation (\ref{hamiltonian2}). There results an
inequality of the form $\dot l\ge C_2 l^{n+1}$, where the positive constant
$C_2$ depends only on $C_1$ and $B$. This in turn means that $l^{-n}$ must
tend to zero within a finite time only depending on $C_2$, provided the
solution exists that long. Thus it has been shown under the given 
assumptions that if every solution continues to exist as long as $l$ remains
bounded then every solution will end in a big rip singularity in
finite time. The existence statement is dependent on the choice of matter 
model and will be addressed later.

It will now be shown that the geometry isotropizes in the approach to 
the big rip singularity in these models. This means by definition that
$\sigma_{ij}\sigma^{ij}/H^2$ tends to zero as $t$ tends to
the time $t_*$ where $l$ tends to infinity. Since $Z$ is bounded by a 
constant times $l^{-2}$ as $t\to t_*$ it follows from the definition of
$Z$ that $\sigma_{ij}\sigma^{ij}=O(l^{-2})$ in this limit. On the the other
hand by the modified Hamiltonian constraint it follows that 
$H^2\ge \frac{8\pi}{3}Bl^{-2n}$ and so the quotient of interest decays at
least as fast as $l^{-2n-2}$ as $t\to t_*$. The quotients $\rho/H^2$ and
$R/H^2$ satisfy similar estimates which says that the density becomes
negligible compared to the critical density and the spatial scalar curvature
becomes unimportant in the approach to the big rip singularity. 

It is possible to refine these statements so as to obtain more detailed
information about the asymptotics of the geometry in the limit $t\to t_*$.
This is analogous to the refinement of Wald's theorem proved by Lee 
\cite{lee}. The first step is to determine the leading order 
asymptotics of $\tr k$. From the modified Hamiltonian constraint and
the estimates already obtained it follows that
\begin{equation}
(\tr k)^2=24\pi f(\rho)(1+o(1)).
\end{equation} 
Consider now the right hand side of equation (\ref{trkevol}).
\begin{equation}
R+(\tr k)^2+4\pi\tr S-12\pi\rho=(\tr k)^2 (1+o(1)).
\end{equation} 
For a power law Cardassian term $\rho f'(\rho)=nf(\rho)$. In the case of a 
perfect fluid with equation of state $p=(\gamma-1)\rho$ the explicit 
contribution containing $\tr S$ coming from (\ref{decomp})
adds to $4\pi\rho f'(\rho)$ to give 
$4\pi\gamma\rho f'(\rho)=4\pi n\gamma f(\rho)$. Hence
\begin{equation}
12\pi(-2f(\rho)+(\rho+\tr S) f'(\rho))=12\pi(n\gamma-2)f(\rho)
=\frac12 (n\gamma-2)(\tr k)^2(1+o(1)). 
\end{equation}
With the information available concerning $Z$ the estimates previously 
obtained for the terms containing $\nabla_i j^i$ and $k^m{}_lS^l{}_m$
can be improved. Putting all these facts together shows that 
\begin{equation}
\frac{d}{dt}(\tr k)=\frac{n\gamma}{2}(\tr k)^2 (1+o(1)).
\end{equation}
It follows that $\tr k=\frac{2}{n\gamma}(t_*-t)^{-1}(1+o(1))$ and that
$l=l_0 (t_*-t)^{\frac{2}{3n\gamma}}$ for a constant $l_0$. Thus the 
exact leading order asymptotics for $\tr k$ and $l$ have been obtained.

With $\tr k$ under control the leading order asymptotics of the
metric can also be determined. The starting point is the equation
$\frac{d}{dt}(g_{ij})=-2g_{il}k^l{}_j$. It follows from this that
$\frac{d}{dt}(l^{-2} g_{ij})=-2l^{-2}g_{il}\sigma^l{}_j$. By using 
some linear algebra as in \cite{rendall94} it is possible to obtain an 
estimate of the form
\begin{equation} 
\|l^{-2}g (t)\|\le\|l^{-2}g (t_0)\|+\int_{t_0}^t \|\sigma(s)\|
\|l^{-2}g (s)\| ds.
\end{equation}
Here $t_0$ is some fixed time and the norms are the matrix norms of
the matrices of components of the corresponding geometric objects.
The norm of $\sigma$ can be estimated in terms of $\sigma_{ij}\sigma^{ij}$
for which a decay estimate in the approach to $t=t_*$ has already been
obtained. There result bounds for the norm of $l^{-2}g$ and for the 
components $l^{-2} g_{ij}$. An analogous argument can be applied to
bound the components $l^2 g^{ij}$. Thus the metric $l^{-2} g_{ij}$
is uniformly equivalent to the flat metric with components $\delta_{ij}$.
Using the evolution equation for $l^{-2}g_{ij}$ again shows that this
quantity converges to a limit $g^*_{ij}$ as $t\to t_*$. Its inverse
also converges so that $g^*_{ij}$ is non-degenerate. Finally the
leading term in the expansion of the metric can be derived. The results 
which have been proved assuming a continuation criterion will now be summed 
up in a theorem.

\noindent
{\bf Theorem} Consider a spatially homogeneous solution of Bianchi type 
I-VIII of the Cardassian equations (\ref{hamiltonian}) and (\ref{kevol}) in 
the case that $f(\rho)=B\rho^n$ with $B>0$ and $n<0$. Suppose that the matter 
is described by a perfect fluid with equation of state $p=(\gamma-1)\rho$, 
$1\le\gamma\le 2$. If the solution is expanding at some time then the 
determinant of the metric tends to infinity at a finite time $t_*$ in the 
future. For $t\to t_*$
\begin{equation}
g_{ij}(t)=(t_*-t)^{\frac{4}{3n\gamma}}g^*_{ij}+o((t_*-t)^{\frac{4}{3n\gamma}})
\end{equation}
and $\tr k\to\frac{2}{n\gamma}$.

Note how similar this result is to that in the homogeneous case. As in that
case more detailed asymptotics can be obtained if desired \cite{berndt}.
To complete the proof of this theorem a continuation criterion must be proved. 
Is is contained in the following Lemma.

\noindent
{\bf Lemma} Under the assumptions of the theorem if the solution exists 
on a time interval $[t_0,t_1)$ for some finite $t_1$ and the determinant of 
the metric remains bounded as $t\to t_1$ then the solution can be extended
to an interval $[t_0,t_2)$  for some $t_2>t_1$. 

\noindent
{\bf Proof} A key fact in proving this lemma is the boundedness of
$Z$. The energy density satisfies the evolution equation
\begin{equation}
\frac{d\rho}{dt}=(\tr k)(\rho+p)-\nabla_i j^i+\sigma^i{}_jk^j{}_i
=\tr k\gamma\rho+O(\rho).
\end{equation}
here it has been used that $Z$ is bounded in order to bound the term
containing $\sigma^i{}_j$. This implies that
\begin{equation}
\frac{d}{dt}(\log (\rho l^{3\gamma}))=O(1).
\end{equation}
It can be concluded that since $l$ is bounded on the given interval the
same is true of $\rho^{-1}$. Using the fact that $Z$ is bounded again 
shows that $\tr k$ is bounded. It is already known that $l^{-1}$ and
$\rho$ are bounded. The bounds for $\tr k$ and $\sigma_{ij}\sigma^{ij}$
can be used to bound $g_{ij}$, $g^{ij}$ and $k_{ij}$ by a method which
was already used above. It follows from the boundedness of $\tr k$ that
$(\tr k)^{-1}$ is also bounded, as was shown following (\ref{gronwall}). 
From this point the rest of the proof 
proceeds essentially as in section 4 of \cite{rendall95}. The solution
remains in a compact subset of the region where the coefficients of 
the evolution equations for the second fundamental form are regular as a 
consequence of the fact that $\tr k$ cannot approach zero. 

The analysis of the lemma can be extended to other reasonable matter models.
This includes mixtures of non-interacting fluids with linear equations
of state. It also includes the case of collisionless matter satisfying the 
Vlasov equation. This provides a model of the distribution of galaxies
which includes velocity dispersion. The only new element which comes in
is the need for a suitable local existence theorem with continuation
criterion. This can be proved in the same way for these matter models 
as in the case of a single fluid already discussed. The proof follows
the arguments in section 4 of \cite{rendall95} as before and, in the
case of collisionless matter, those of section 2 of \cite{rendall94}.
This implies a partial analogue of the theorem for the other matter models.
The statements about the occurrence of a big rip singularity and the
isotropization of the geometry as it is approached go through without
change. More precise information concerning the leading terms in the
asymptotics depend on the details of the matter model chosen. For it
is necessary to know the leading order behaviour of the pressure as 
a function of the energy density. In the case of a mixture of fluids the 
fluid with the smallest value of $\gamma$ will dominate. In the case of 
collisionless matter the leading order behaviour will be like that of 
dust since for collisionless matter $\tr S/\rho\to 0$ when the scale factor 
tends to infinity.

\section{Discussion}

In this paper the Cardassian models in cosmology were extended to a
large class of geometries which are homogeneous but not necessarily 
isotropic. It was shown that for Bianchi types I-VIII, matter described
by a perfect fluid with linear equation of state and Cardassian
terms of the form $B\rho^n$ with $B>0$ and $n<0$ the property of
isotropic models that all solutions develop a big rip singularity
extends to all the homogeneous models. Moreover the homogeneous models
approach isotropic and spatially flat models in a strong sense as the
big rip singularity is approached.

At the same time it was pointed out that there is no obvious way of
extending further to the inhomogeneous case so that it is not really
possible to call the Cardassian models a theory of gravitation. 
This statement does depend on the point of view that reinterpreting
the homogeneous and isotropic models in terms of an exotic fluid and
passing to the inhomogeneous case in the context of the Einstein equations
coupled to that fluid is not a satisfactory solution to the problem.
Arguments supporting this point of view were presented. It is difficult
to rule out that there is some natural general inhomogeneous version of 
the Cardassian models but the authors have failed to find one.

At a time when new models designed to address the question of the
nature of the cause of accelerated expansion are proliferating, some of them 
differing from general relativity, it would be good to pose the question at an
early stage of the history of such a model whether there is indeed
a general theory behind it. There are cases where this is clear, as
with the $f(R)$ theories which ar presently very popular or other higher 
order theories such as those with Lagrangian $R_{\alpha\beta}R^{\alpha\beta}$. 
This includes the fact that the field equations for these theories are known 
to be well posed \cite{noakes}. The example of the Cardassian models shows 
that things are not always so simple. It may be that the range of 
applicability of a proposed model does not go beyond the homogeneous case.

Big rip singularities can occur for other models which have been 
introduced to explain accelerated cosmological expansion. This applies in
particular to general relativity coupled to certain types of exotic
matter. It may well be that the isotropization phenomenon exhibited in
this paper for Cardassian models also occurs in the approach to big rip 
singularities produced by other mechanisms. This remains to be investigated.

\section{Appendix: motion in a potential}

Let $V$ be a smooth function defined on an interval $(a_-,a_+)$ of real
numbers where $a_-$ and $a_+$ may be finite or infinite. To exclude 
trivialities suppose that $V$ is not constant. Consider the 
equation $\ddot a=V'(a)$. The energy $\frac{\dot a^2}2+V(a)$ is independent 
of time. Corresponding to initial data at some time $t_0$ for a solution of 
this equation there is a solution on a maximal interval $(t_-,t_+)$ beyond 
which it cannot be extended smoothly. Here $t_-$ and $t_+$ may be finite or 
infinite. If one of them is finite then as $t$ tends to this endpoint
$a(t)$ must come arbitrarily close to $a_-$ or $a_+$. For if not conservation
of energy implies that $\dot a$ remains bounded. Then $a$ and $\dot a$ 
remain in a compact subset of the admissible values where the coefficients
of the equation are regular and by a standard result on ordinary differential
equations \cite{hartman} the solution can be extended, contradicting the 
original assumption. 

Let $a(t)$ be a solution with initial data given by $a(t_0)=a_0$ and
$\dot a(t_0)=a_1$. Its energy is $E=\frac12 a_1^2+V(a_0)$. Let 
$b_-=\inf\{a:V(a)<E\ {\rm on}\ (b_-,a_0)\}$ if the set over which the infimum 
is to be taken is non-empty and $b_-=a_0$ otherwise. Correspondingly, let 
$b_+=\sup\{a:V(a)<E\ {\rm on}\ (a_0,b_+)\}$ if the set over which the supremum 
is to be taken is non-empty and $b_+=a_0$ otherwise. If $b_-=b_+$ then 
$V'(a_0)=0$ and the solution is time independent. Otherwise the solution 
defines an interval $[b_-,b_+]$ of positive length. Suppose that $b_->a_-$
and that the solution reaches the point $b_-$ at a finite time $t_1>t_0$. 
Then $\dot a (t_1)=0$ and $V'(b_-)\le 0$. If $V'(b_-)$ were zero the solution
could only be the time independent solution at $b_-$. Thus $V'(b_-)<0$ and 
$a$ is greater than $b_-$ at times slightly after $t_1$. In summary, if the
solution reaches a point $b_-$ in the interval $(a_-,a_+)$ after finite
time it must turn around and start increasing. The same argument applies to 
the evolution backwards in time. An analogous statement can be obtained 
with $b_-$ replaced by $b_+$ by a similar argument. A solution behaves in
a monotone way as long as it is strictly between $b_-$ and $b_+$ since 
$\dot a$ can never vanish there. Thus if in a given solution $\dot a(t_0)>0$ 
is $\dot a$ will increase for all $t\ge t_0$ unless $b_+\in (a_-,a_+)$. In 
the  latter case $\dot a$ will change sign and $a$ will start to decrease.
If $\dot a(t_0)<0$ then $\dot a (t)$ will decrease for all $t\ge t_0$ unless
$b_-\in (a_-,a_+)$. Similar statements apply to the past time direction.
If both $b_-$ and $b_+$ are in $(a_-,a_+)$ then there are two times $t_1$ and 
$t_2$ at which the solution reaches $b_-$. At these times the initial data 
are identical and so the solution is periodic, oscillating between the two 
endpoints $b_-$ and $b_+$. Thus the following cases are possible. The 
solution is time independent (type 1). The solution is periodic (type 2).
The solution is such that $\dot a$ changes sign exactly once (type 3).
The solution is such that the sign of $\dot a$ is constant (type 4). 
Which of these types occur is easily decided in terms of the position of
$b_-$ and $b_+$. In types 1 and 2 this already determines the essential 
features of the dynamics. In types 3 and 4 it remains to decide the 
following. How does the solution behave during the monotone approach to
the point $a_-$ or $a_+$? It can be assumed without loss of generality that
$\dot a(t)>0$ in the time interval of interest and approaches $a_+$.
It should be remembered that $a_+$ may be finite or infinite and that,
independently of this, $t_+$ may be finite or infinite. For the case 
being considered here to occur it must be the case that $V(a)<E$ 
close to $a_+$. By conservation of energy $\dot a=\sqrt{2(E-2V(a))}$.
Thus what happens depends on the details of the limiting behaviour
of $V$ as $a\to a_+$ and its relation to $E$. In the examples of relevance 
in this paper it is always the case that $a_-=0$ and $a_+=\infty$.

\end{document}